\documentclass[pra,twocolumn,nofootinbib,longbibliography]{revtex4-1}
\pdfoutput=1
\usepackage{hyperref}
\usepackage{comment}
\usepackage{amsmath}
\usepackage{amssymb}
\usepackage{mathtools}
\newcommand{\eqnref}[1]{Eq.~(\ref{#1})}
\begin{document}
\title{Collisionless dynamics of general non-Fermi liquids from hydrodynamics of emergent conserved quantities}
\author{Dominic V. Else}
\affiliation{Perimeter Institute for Theoretical Physics, Waterloo, ON N2L 2Y5, Canada}
\begin{abstract}
    Given the considerable theoretical challenges in understanding strongly coupled metals and non-Fermi liquids, it is valuable to have a framework to understand properties of metals that are \emph{universal}, in the sense that they must hold in any metal. It has previously been argued that an infinite-dimensional emergent symmetry group is such a property, at least for clean, compressible metals. In this paper, we will show that such an emergent symmetry group has very strong implications for the dynamics of the metal. Specifically, we show that consideration of the hydrodynamics of the associated infinitely many emergent conserved quantities automatically recovers the collisionless Boltzmann equation that governs the dynamics of a Fermi liquid. Therefore, the hydrodynamic prediction is that in the low-temperature, collisionless regime where the emergent conservation laws hold, the dynamics and response to external fields of a general spinless metal will be \emph{identical} to a Fermi liquid. We discuss some potential limitations to this general statement, including the possibility of non-hydrodynamic modes. We also report some interesting differences in the case of spinful metals.
\end{abstract}
\maketitle
A metal is one of the basic categories of phases of matter that a many-body system of electrons can be in. The low-energy physics of metals can exhibit many rich phenomena. Fermi liquid theory \cite{Baym,Shankar_9307} is a highly successful effective field theory that captures the low-energy physics of many metallic materials in terms of the dynamics of quasiparticles. Nevertheless, Fermi liquid theory has turned out not to be an adequate description of so-called ``non-Fermi liquid'' metals that have been observed in various classes of materials.

Theoretical progress in understanding non-Fermi liquids has been slow despite intense interest. A major focus has been in studying models of fermions coupled to a critically fluctuating bosons \cite{Hertz__1976,Millis__1993,Lohneysen_0606}. In two spatial dimensions, by deforming the theory in certain ways (for example by introducing $N$ fermion flavors in a particular way), certain limits have been found in which perturbatively controlled calculations may be possible \cite{Mross_1003,Dalidovich_1307,Lee_1703,Esterlis_2103}. However, the physics in these limits may be qualitatively different from the more physical undeformed models \cite{Shi_2204,Shi_2208}. Moreover, there is no reason to believe that these ``Hertz-Millis'' type models exhaust the possibilities of non-Fermi liquids or are necessarily the appropriate description of the non-Fermi liquids seen in experiment.

A dramatically different way to think about the problem is to ask which properties of metals are \emph{universal}\footnote{Here we are not using ``universal'' in the common sense of properties captured by a single renormalization-group (RG) fixed point; rather we are talking about properties that are common to \emph{many different} RG fixed points of metals. Perhaps a better term would be ``multiversal''.} in the sense that they are common to all (or at least most) metals, even strongly interacting ones? Some progress in addressing this question has been made in Refs.~\cite{Else_2007,Else_2010,Else_2106}. The main idea is that in clean systems, i.e.\ systems with at least a lattice translation symmetry, one can use \emph{compressibility} as a proxy for metallicity, where ``compressibility'' here is defined as the ability of the microscopic filling, i.e.\ the number of electrons per unit cell, to be tuned continuously, possibly in tandem with other microscopic parameters, while remaining in the same phase. Compressibility is a feature of all known theories of metals, and to our knowledge is compatible with all experimental observations of metals as well. (One exception is the so-called ``composite Fermi liquids'' that occur in the context of the fractional quantum Hall effect \cite{Halperin__1993,Son_1502}, though whether these should even be called metals is debatable since the longitudinal conductivity is zero at zero temperature in a clean system).

Ref.~\cite{Else_2007} showed that compressibility in fact has dramatic consequences for the low-energy theory: in spatial dimension $d > 1$, the emergent symmetry group must either include a so-called higher-form symmetry (which is what happens in superfluids, another example of a compressible phase), or else it must be infinite-dimensional. The latter possibility is realized by Fermi liquid theory: in which at temperatures low enough that quasiparticle scattering can be disregarded, the charge at \emph{each} point on the Fermi surface is conserved. 

These considerations led us in Ref.~\cite{Else_2007} to introduce the concept of an \emph{ersatz Fermi liquid}, which is a system with the same emergent symmetry group (and hence, the same structure of emergent conserved quantities) as a Fermi liquid. It is an important open question to determine whether ersatz Fermi liquids and variations thereof capture all possible compressible metals, or whether there are other fundamentally different possibilities. Nevertheless, ersatz Fermi liquids at least represent a large class of non-Fermi liquids (including, in particular, Hertz-Millis type models). We will restrict ourselves to ersatz Fermi liquids in this paper.

Ref.~\cite{Else_2007} demonstrated that a number of properties of Fermi liquids can, in fact, be deduced strictly from the emergent symmetry, and therefore constitute \emph{universal} properties of any ersatz Fermi liquid. These properties included, for example, Luttinger's theorem. In the present work, we will considerably extend these results by turning our attention to the dynamical properties of the system; that is, the oscillation modes at low frequencies and wavelengths, as well as the responses to external electric fields, such as the conductivity tensor $\sigma(\omega, \mathbf{q})$. These are the properties that in Fermi liquid theory can be analyzed through the kinetic equation of the quasiparticles.

It should not be surprising that the infinitely many emergent conservation laws will have important implications for the dynamical properties of the system. In this paper, we show that the constraint is actually maximally strong: to the extent that the dynamics of the system is governed by the hydrodynamics of the conserved quantities (an assumption we will re-examine towards the end), the linearized dynamics of a general ersatz Fermi liquid is in fact \emph{identical} to that of a Fermi liquid with suitable Fermi velocity and Landau interaction parameters. Thus, all the dynamical features of Fermi liquids such as the collective ``zero sound'' mode and the particle-hole continuum will carry over. This work represents a clear demonstration of the power of the compressibility and emergent symmetry concepts in understanding the dynamics of non-Fermi liquids.

Finally, let us remark that this paper can in many ways be viewed as a sequel to Ref.~\cite{Delacretaz_1908}. In that paper it was shown that the Goldstone modes of a superfluid can in fact be derived solely from the emergent symmetry [in that case, in $d$ spatial dimensions it is a $(d-1)$-form symmetry] and its mixed anomaly with the charge $0$-form symmetry; as a consequence, they are present even in systems where the $\mathrm{U}(1)$ charge conservation symmetry is not actually spontaneously broken (which we could refer to as ``ersatz superfluids''). In this paper we are implementing an analogous program for ersatz Fermi liquids. However, as the emergent symmetry group in this case is much larger, the dynamical modes that one can obtain are richer, as we will see.

The outline of the remainder of this paper is as follows. In Section \ref{sec:fl_dynamics} we will review the dynamics of a Fermi liquid from the Boltzmann equation. In Section \ref{sec:ersatz}, we will review the concept of an ersatz Fermi liquid. In Section \ref{sec:ersatz_dynamics} we will state the precise result for the equation of motion that we are going to establish in a general ersatz Fermi liquid, and then in Section \ref{sec:zeroth_hydro} we will derive this result from the hydrodynamics of the emergent conserved quantities. In Section \ref{sec:qbe} we will compare with results obtained from the Quantum Boltzmann Equation approach, and reveal how non-hydrodynamic modes can arise in certain regions of $(\omega,\mathbf{q})$ space. In Section \ref{sec:effect_of_criticality} we will discuss how the solutions of the hydrodynamic equations of motion may have different qualitative character at quantum critical points due to special values of the parameters. In Section \ref{sec:extensions} we will discuss possible extensions of our results to spinful systems and to the response to magnetic fields. In Section \ref{sec:discussion} we will discuss limitations of our results and future directions.

    \section{Review: dynamics of a Fermi liquid}

    \label{sec:fl_dynamics}

As is well known, in (spinless) Fermi liquid theory, in the collisionless regime where quasiparticle scattering can be disregarded, the quasiparticle distribution function $f(\mathbf{x},\mathbf{k},t)$ obeys the collisionless Boltzmann equation
\begin{equation}
    \label{eq:boltzmann}
    \frac{\partial f}{\partial t} - \left( \mathbf{E} + \frac{\partial \epsilon}{\partial \mathbf{x}} \right) \cdot \frac{\partial f}{\partial \mathbf{k}} + \frac{\partial \epsilon}{\partial \mathbf{k}}\cdot \frac{\partial f}{\partial \mathbf{x}} = 0
\end{equation}
where
\begin{equation}
    \label{eq:qp_energy}
\epsilon(\mathbf{k},\mathbf{x},t) = \epsilon_0(\mathbf{k}) + \int F(\mathbf{k},\mathbf{k}') \delta f(\mathbf{k}',\mathbf{x},t) d^d \mathbf{k}'
\end{equation}
is the energy of a single quasiparticle. Here $\epsilon_0(\mathbf{k})$ is the equilibrium value of the quasiparticle energy, $F(\mathbf{k},\mathbf{k}')$ the Landau interaction, and $\delta f = f - f_0$ is the deviation from the  distribution function in the ground state. For generality, we have included an electric field $\mathbf{E}$ which could depend on space and time. In the majority of the paper we will set the \emph{magnetic} field to zero (however, we will make some comments on magnetic fields towards the end).

We work in general spatial dimension $d$. We parameterize the ground state Fermi surface in momentum space by $\mathbf{k}_F(\theta)$, where $\theta$ is a parameter that lives in some closed $(d-1)$-dimensional manifold $\mathcal{S}$ with the appropriate topology [for example a $(d-1)$-sphere.] The long-wavelength, low-frequency dynamics of the system can be described in terms of a perturbed Fermi surface $\mathbf{K}(\theta)$ that differs by a small amount from $\mathbf{k}_F(\theta)$ and could depend on space and time. In particular, at zero temperature we can set $f(\mathbf{x},\mathbf{k},t)$ to be $1/(2\pi)^d$ when $\mathbf{k}$ is inside the perturbed Fermi surface at $(\mathbf{x},t)$ and zero outside. If we substitute into \eqnref{eq:boltzmann} and linearize in the perturbation (treating the external electric field to linear order as well), we obtain a linear equation of motion for the Fermi surface.

The equation of motion will involve the component of the perturbation perpendicular to the Fermi surface, namely $\hat{\mathbf{w}}(\theta) \cdot (\mathbf{K}(\theta) - \mathbf{k}_F(\theta))$, where $\hat{\mathbf{w}}(\theta)$ is a unit vector normal to the Fermi surface. [The components of $\mathbf{K}(\theta) - \mathbf{k}_F(\theta)$ parallel to the Fermi surface can be eliminated by a time- and space-dependent reparameterization of $\theta$ and have no physical content.] 
For reasons that will become clear later, we will prefer to introduce a non-unit vector $\mathbf{w}(\theta)$ that is normal to the Fermi surface and define
\begin{equation}
    n(\theta) = \frac{1}{(2\pi)^d} \mathbf{w}(\theta) \cdot [\mathbf{K}(\theta) - \mathbf{k}_F(\theta)].
\end{equation}
The idea is that $\int n(\theta) d\theta$ will be equal to the total excess charge density. Furthermore, $\int_{\Sigma} n(\theta) d\theta$, where the integral is restricted to a region $\Sigma$ of the Fermi surface, will give the contribution to the excess charge density from that portion of the Fermi surface. Here $\int d\theta$ denotes integration with respect to some arbitrarily chosen volume form on $\mathcal{S}$. Integrals will be assumed to be over the whole manifold $\mathcal{S}$ unless otherwise stated.

In order to define $\mathbf{w}(\theta)$, suppose that in some infinitesimal neighborhood in $\theta$ space, we choose some coordinate system $\theta_1, \cdots, \theta_{d-1}$, normalized such that within this local neighborhood, the integration measure $\int d\theta$ coincides with the usual integration measure in $\mathbb{R}^d$. Then within this neighborhood we can define
\begin{equation}
    \label{eq:w_defn}
  w^i(\theta) = \epsilon^{i j_1 \cdots j_{d-1}} \partial_{\theta_1} (k_F)_{j_1}(\theta) \cdots \partial_{\theta_{d-1}} (k_F)_{j_{d-1}}(\theta).
\end{equation}
where $\epsilon$ denotes the Levi-Civita symbol, and we use the repeated index summation convention.

In terms of $n(\theta)$, the linearized equation of motion can be written as
\begin{multline}
    \label{eq:lineq}
    \frac{\partial n(\theta)}{\partial t} + \mathbf{v}_F(\theta) \cdot \nabla n(\theta) - \frac{1}{(2\pi)^d} \mathbf{w}(\theta) \cdot \mathbf{E}
    \\ + \frac{1}{(2\pi)^d} \int d\theta' F(\theta, \theta') \mathbf{w}(\theta) \cdot \nabla n(\theta') d\theta' 
    = 0
\end{multline}
where $\mathbf{v}_F(\theta) := \frac{\partial \epsilon_0}{\partial \mathbf{k}} \bigr|_{\mathbf{k} = \mathbf{k}_F(\theta)}$ is the Fermi velocity, and $F(\theta,\theta') := F(\mathbf{k}_F(\theta), \mathbf{k}_F(\theta'))$ is the Landau integration evaluated at the Fermi surface. [Note that by the definition of the Fermi surface, $\mathbf{v}_F(\theta)$ must be normal to the Fermi surface and hence parallel to $\mathbf{w}(\theta)$.] Since we are only considering the linearized equation of motion, it is not necessary to take into account the variation of the Fermi wave vector from the equilibrium value $\mathbf{k}_F(\theta)$ in defining $F(\theta,\theta')$, $\mathbf{v}_F(\theta)$ and $\mathbf{w}(\theta)$. We remark that while above we talked about zero temperature, \eqnref{eq:lineq} actually holds at nonzero temperature as well, up to leading order in $T$ (and ignoring collisions).

\begin{figure}
%
%
%
%
    \includegraphics{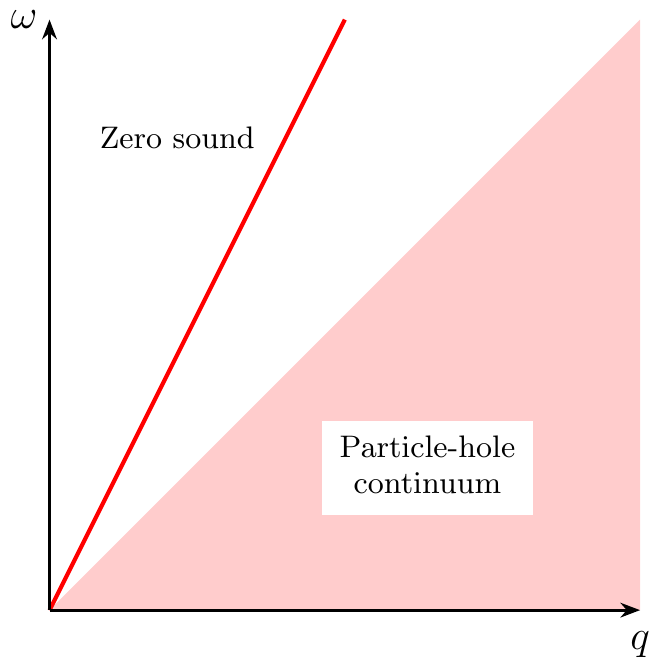}
    \caption{\label{fig:fl_spectrum} The oscillation mode spectrum for a Fermi liquid (the zero sound mode may or may not be present depending on the value of the Landau parameters).}
\end{figure}

All of the low-frequency, long wavelength dynamical properties of a Fermi liquid at zero temperature (except those relating to magnetic fields) can be determined by solving \eqnref{eq:lineq}. For example, one can compute the conductivity tensor $\sigma^{ij}(\omega, \mathbf{q})$ and the density-density response function $G_{nn}(\omega,\mathbf{q})$. Moreover, by solving \eqnref{eq:lineq} in the absence of any external fields, one finds the oscillation modes of the system at given frequency $\omega$ and wave-vector $\mathbf{q}$. What one generally finds, as shown in Figure \ref{fig:fl_spectrum}, is a continuum of modes referred to as the ``particle-hole continuum''. (Depending on the specific parameter values, there can be an additional collective mode called ``zero sound''). The particle-hole continuum exists at $\omega$ and $\mathbf{q}$ such that there exists a point $\theta$ on the Fermi surface with $\omega = \mathbf{v}(\theta) \cdot \mathbf{q}$. (In particular, for an isotropic system it exists for $|\omega|/q < v_F$).

\section{Review: ersatz Fermi liquids}
\label{sec:ersatz}
    In Fermi liquid theory, the total charge $N(\theta) = \int n(\theta,\mathbf{x}) d^d \mathbf{x}$ associated with \emph{each} point on the Fermi surface is a conserved quantity, reflecting the fact that quasiparticles do not scatter. [Specifically, this is a property of the RG fixed-point theory. Irrelevant operators not included in the fixed-point theory would lead to quasiparticle scattering at nonzero temperature or frequency.] In other words, we can think of the $N(\theta)$'s as the generators of an infinite-dimensional \emph{emergent} symmetry group, which in $d=2$ be identified as the so-called ``loop group'' $\mathrm{LU}(1)$ \cite{Else_2007}. For general $d$, the emergent symmetry group is $\mathrm{L}_{\mathcal{S}}\mathrm{U}(1)$, the group whose elements are smooth maps into $\mathrm{U}(1)$ from the closed manifold $\mathcal{S}$ that parameterizes the Fermi surface.

An ersatz Fermi liquid is defined to be any system that has the same $\mathrm{L}_{\mathcal{S}}\mathrm{U}(1)$ emergent symmetry group, for some $(d-1)$-dimensional closed manifold $\mathcal{S}$. In particular, we will refer to the generators of this group as $N(\theta)$, with $\theta$ living in $\mathcal{S}$, as in a Fermi liquid. Note that, by examining how the microscopic lattice translation group maps into $\mathrm{L}_{\mathcal{S}} \mathrm{U}(1)$, one immediately obtains a concept of ``Fermi surface'' that applies in an arbitrary Fermi liquid \cite{Else_2007}. We will continue to parameterize this Fermi surface by $\mathbf{k}_F(\theta)$ as in a Fermi liquid.

An important property of this emergent symmetry group is its 't Hooft anomaly. In this context, we can think of a 't Hooft anomaly as a non-conservation of charge in response to a background gauge field. An example would be the anomaly of a chiral fermion in (1+1)-D where charge becomes nonconserved in the presence of an applied electric field.
 As the details get somewhat technical, we refer the reader to Ref.~\cite{Else_2007} for a general definition of the 't Hooft anomaly of $\mathrm{L}_{\mathcal{S}} \mathrm{U}(1)$. Here we will simply mention some of its consequences. Among them is that the charge $N(\theta)$ becomes non-conserved when a background electric field is applied, according to\footnote{In Ref.~\cite{Else_2007} there was an additional factor of $q$ in this equation, where $q$ is an integer that reflects how the microscopic $\mathrm{U}(1)$ charge conservation symmetry maps into $\mathrm{L}_{\mathcal{S}} \mathrm{U}(1)$. In Fermi liquid theory, $q$ represents the charge carried by a quasiparticle. In this paper we will just absorb $q$ into the definition of the electric field, as we already did implicitly in Section \ref{sec:fl_dynamics}.}
 \begin{equation}
        \label{eq:conservation_eqn}
    \partial_t n(\theta) + \nabla \cdot \mathbf{j}(\theta) = \frac{m}{(2\pi)^d} \mathbf{w}(\theta) \cdot \mathbf{E},
 \end{equation}
 where $\mathbf{w}(\theta)$ is defined by \eqnref{eq:w_defn}, $n(\theta)$ is the local density of $N(\theta)$, and $\mathbf{j}(\theta)$ is the corresponding current density. We have introduced the anomaly coefficient $m$, which is quantized to be an integer. For single-component Fermi liquid theory, $m = \pm 1$. More generally, if there are $N$ fermion species that share a Fermi surface, then $|m|=N$ in Fermi liquid theory.
 
 Another consequence of the anomaly (though not one we will need to use in the current paper) is Luttinger's theorem, which holds in any ersatz Fermi liquid that has a microscopic lattice translation symmetry and $\mathrm{U}(1)$ charge conservation symmetry\footnote{Here there again should technically be an additional factor of $q$, see the previous footnote.}:
 \begin{equation}
    \frac{m\mathcal{V}_F}{(2\pi)^d} = \nu \quad(\operatorname{mod} 1)
 \end{equation}
 where $\mathcal{V}_F$ is the volume enclosed by the Fermi surface, and $\nu$ is the microscopic filling, i.e.\ the average charge per unit cell.

\section{Dynamics of ersatz Fermi liquids}
\label{sec:ersatz_dynamics}

The emergent conservation laws in an ersatz Fermi liquid have important consequences for the dynamics. In particular, whenever we have conserved quantities, then we can treat the dynamics of the densities of the conserved quantities from the point of view of hydrodynamics, by assuming that for sufficiently low frequency $\omega$ and wave-vector $\mathbf{q}$, the system can be regarded as reaching local thermal equilibrium at any given point in space and time, characterized
solely by the local values of the densities of the conserved quantities (or equivalently, their thermodynamically conjugate variables). One supplements this with constitutive relations that relate the densities and currents of the conserved quantities to the thermodynamically conjugate quantities and their derivatives. One can view these constitutive relations in terms of a derivative expansion, where one truncates the series by retaining only terms with a small number of derivatives.

A familiar example of hydrodynamics is the case where the only conserved quantities are charge/mass, energy, and momentum. This gives the usual hydrodynamics of fluids. If we truncate the derivative expansion to ``zero-th'' order, where one discards all terms in the constitutive relations that involve derivatives, then one obtains the Euler equations of fluid dynamics; going to first-order in the derivative expansion instead gives the Navier-Stokes equations that include viscosity terms. (In general, zero-th order hyrodynamics will always be non-dissipative, while dissipation effects enter at higher orders).

The central result of this paper is that, if we formulate hydrodynamics taking into account the emergent conserved quantities of ersatz Fermi liquids, and if one truncates the derivative expansion at ``zero-th'' order, then one precisely obtains equation of motion for $n(\theta)$ that corresponds to \eqnref{eq:lineq}.

The precise statement that we are going to prove is the following. A thermal equilibrium state can be characterized by the expectation values of the densities of all conserved quantities. Suppose that the conserved quantities of the system are $N(\theta)$, the energy $E$, and potentially some additional ``spectator'' conserved quantities $O_1, O_2, \cdots$ (we will clarify what exactly qualifies as a ``spectator'' quantity later; here we just remark that an example would be total spin in systems of spinful electrons -- see Section \ref{subsec:spinful}).
 We define a thermodynamic variable $\mu(\theta)$ conjugate to $n(\theta)$ according to
\begin{equation}
    \mu(\theta) = \left[\frac{\delta \varepsilon}{\delta n(\theta)}\right]_{s,o}
\end{equation}
where $\varepsilon$ is the energy density, and the notation indicates that the derivative is taken with the entropy density $s$, as well as the densities of all the spectator quantities, held fixed.
Then we define a function $\xi$ according to
\begin{equation}
    \label{eq:xi}
    \xi(\theta,\theta') = \left[\frac{\delta \mu(\theta)}{\delta n(\theta')}\right]_{s,o}.
\end{equation}
We can essentially think of $\xi$ as related to the thermodynamic susceptibilities, although taking the derivative with entropy held fixed is not quite how susceptibilities are normally defined. (However, the Third Law of Thermodynamics ensures that this at zero temperature, derivatives with entropy held fixed are equivalent to derivatives with temperature held fixed, so this distinction only matters at nonzero temperature.) By exchanging the order of derivatives one immediately finds that $\xi$ is symmetric, that is, $\xi(\theta,\theta') = \xi(\theta',\theta)$.

What we will show is that truncating to the zero-th order of the hydrodynamic expansion gives the equation of motion
\begin{multline}
    \label{eq:general_motion}
    \frac{\partial n(\theta)}{\partial t} + \frac{m}{(2\pi)^d} \int d\theta' \xi(\theta, \theta') \mathbf{w}(\theta) \cdot \nabla n(\theta') d\theta' \\= \frac{m}{(2\pi)^d} \mathbf{w}(\theta) \cdot \mathbf{E}.
\end{multline}
In particular, if one sets $m=1$ and substitutes the value of $\xi$ in single-component Fermi liquid theory, which can be shown (see Appendix \ref{appendix:fl_susceptibility}) to be
\begin{equation}
    \label{eq:xi_landau}
    \xi(\theta,\theta') = (2\pi)^d \frac{v_F(\theta)}{|\mathbf{w}(\theta)| }\delta(\theta - \theta') + F(\theta,\theta'),
\end{equation}
then one recovers \eqnref{eq:lineq}. However, our derivation is much more general and applies to any ersatz Fermi liquid.

It seems reasonable to assume that in a general ersatz Fermi liquid, $\xi(\theta,\theta')$ will be the sum of a non-singular contribution and a delta function contact term.  In that case we can take \eqnref{eq:xi_landau} to be the \emph{definition} of an effective ``Fermi velocity'' $v_F(\theta)$ and ``Landau interaction'' $F(\theta,\theta')$, such that the Fermi-liquid-like equation of motion \eqnref{eq:lineq} is precisely satisfied.

Let us caution that the above terminology may perhaps lead to some confusion in the case of metallic quantum critical points. Suppose that, as one approaches the quantum critical point by tuning some parameter of the Hamiltonian, for any deviation from the quantum critical point, the system remains a Fermi liquid at the lowest energy scales. Such a Fermi liquid can be characterized by the ``effective mass'' (or equivalently, the Fermi velocity) and the Landau interactions. Generally one expects the effective mass to diverge, and hence the Fermi velocity to go to zero, as one approaches the quantum critical point. However, there is generally also a contribution to the Landau interaction $F(\theta,\theta')$ that becomes increasingly sharply peaked near $\theta \approx \theta'$ as one approaches the critical point. At the critical point, this contribution becomes a delta function \cite{Stern_9502} and hence would be re-interpreted as a nonzero contribution to the Fermi velocity, rather than a Landau interaction, according to the above definition. This may not always agree with terminology used in previous literature.

\section{Zero-th order hydrodynamics in an ersatz Fermi liquid}
\label{sec:zeroth_hydro}
We imagine performing a derivative expansion on the constitutive relations. In the present work, we will only retain the zero-th order of this expansion; that is, the terms that do not involve any derivatives at all. These terms just involve thermodynamic susceptibilities. Thus, we have, for example, to linear order in the perturbation about the initial state:
\begin{align}
    \label{eq:delta_mu}
    \delta \mu(\theta) &= \int \xi(\theta,\theta') \delta n(\theta') d\theta' + u(\theta) \delta s + \sum_a \gamma^a \delta o_a
\end{align}
where $s$ is the entropy density, and the $o_a$'s are the densities of the spectator conserved quantities. Here $\xi$ was defined by \eqnref{eq:xi}, $\gamma^a = \left[\frac{\partial \mu(\theta)}{\partial o_a}\right]_{n,s}$, and
\begin{equation}
u(\theta) = \left[\frac{\partial \mu(\theta)}{\partial s}\right]_{n,o} = \left[\frac{\delta T}{\delta n(\theta)}\right]_{s.o}.
\end{equation}
The Third Law of Thermodynamics implies that $u(\theta) \to 0$ at zero temperature. Therefore, we will for the moment just set $u(\theta) = 0$, though we will return to the nonzero temperature case later. We will also temporarily assume that there are no spectator quantities so the last term in \eqnref{eq:delta_mu} goes away.

Now we need to consider the constitutive relations for the currents. A crucial point is that the zero-th order constitutive relation for $\mathbf{j}(\theta)$  is actually completely fixed by general considerations. Since we are dropping all terms in the constitutive relation involving spatial derivatives, it is sufficient to know the expectation value of the current in a thermal equilibrium state. Naively one might think that this is zero due to Bloch's theorem \cite{Yamamoto_1502,Watanabe_1904} but as described in Ref.~\cite{Else_2106} the Bloch's theorem result must be modified in cases where the conserved quantities have an anomaly. In particular, in the present case the arguments of Ref.~\cite{Else_2106} show that the expectation value of the current in an equilibrium state is given by
\begin{equation}
    \label{eq:jtheta_bloch}
    \mathbf{j}(\theta) = \frac{m}{(2\pi)^d} \mathbf{w}(\theta) \mu(\theta).
\end{equation}
 \eqnref{eq:delta_mu} and \eqnref{eq:jtheta_bloch}, combined with the conservation equation \eqnref{eq:conservation_eqn}, immediately implies \eqnref{eq:general_motion}.

Let us now explain how the arguments get modified if there are additional ``spectator'' conserved quantities $O_1, \cdots, O_n$. What we mean by ``spectator'' is that these quantities do not have any mixed anomaly with each other or with $N(\theta)$. Therefore Bloch's theorem will imply that the corresponding currents are zero in an equilibrium state, and therefore can be set to zero in the zero-th order hydrodynamics that we are considering. From this it follows that in studying dynamics we can simply set the perturbation $\delta o^a$ of the densities of these quantities to zero. The absence of a mixed anomaly also implies that there is no correction to \eqnref{eq:jtheta_bloch}. The derivation then proceeds as before, and we again obtain \eqnref{eq:general_motion}.

Finally, let us consider the case of nonzero temperature, in which one cannot set $u(\theta) = 0$ in \eqnref{eq:delta_mu}. We then need to take into account the energy transport. One might think that the energy current in an equilibrium state must be zero -- in fact, this is what is proved for lattice models in Ref.~\cite{Kapustin_1904}. The problem is that the state with $\mu(\theta)$ not a constant function of $\theta$ is not an equilibrium state of a Hamiltonian that can be defined at the lattice level. We give (not entirely rigorous) arguments in Appendix \ref{sec:energy_currents} that the correct value of the energy current in the thermal equilibrium state is
\begin{equation}
    \label{eq:energy_current}
    \mathbf{j}^\varepsilon = \int \mu(\theta) \mathbf{j}(\theta) d\theta,
\end{equation}
Hence, the energy conservation equation gives for the energy density
\begin{align}
    \partial_t \varepsilon &= \mathbf{E} \cdot \mathbf{j} - \nabla \cdot \mathbf{j}^{\varepsilon} \\
    &= \int  \left[\mathbf{E} \cdot \left( \mathbf{j}(\theta) - \frac{m}{(2\pi)^d} \mu(\theta) \mathbf{w}(\theta)\right)+ \mu(\theta) \partial_t n(\theta) \right], \\
    &= \int \mu(\theta) \partial_t n(\theta) d\theta,
\end{align}
where in the second line we invoked \eqnref{eq:energy_current} and the conservation equation \eqnref{eq:conservation_eqn}, and in the third line we invoked \eqnref{eq:jtheta_bloch}. Hence, for the entropy density, we have from a thermodynamic identity that
\begin{align}
    \partial_t s &= \frac{1}{T} \left( \partial_t \varepsilon - \int \mu(\theta) \partial_t n(\theta) d\theta \right) \\ &= 0.
    \label{eq:s0}
\end{align}
From \eqnref{eq:s0} we see that we can set $\delta s = 0$ in \eqnref{eq:delta_mu}, and the rest of the derivation proceeds as before. As a side note we remark that \eqnref{eq:s0} shows that there is no entropy production; in other words, at the level of zero-th order hydrodynamics the dynamics is completely dissipationless.

\section{Comparing with the ``Quantum Boltzmann equation'' formalism: non-hydrodynamic modes}
\label{sec:qbe}
In this section we will compare the results of this paper with those obtained from the ``Quantum Boltzmann equation'' (QBE) formalism \cite{Kim_9504}, which uses non-equilibrium Green's function methods to derive a Boltzmann equation in particular models of fermions coupled to a fluctuating boson. This approach requires the form of the boson and fermion self-energy as input, so it can only be applied in a theory in which these quantities can be computed in a controlled way. The original calculations of Ref.~\cite{Kim_9504} invoked the ``random-phase approximation'' (RPA), but it is now understood \cite{Lee_0905} that this does not represent a controlled approximation even in the limit where the number of fermion species $N_f$ is taken to infinity.

On the other hand, we can apply the QBE in the particular case of the large-$N$ limit of the ``random-flavor'' model discussed in Refs.~\cite{Esterlis_1906,Aldape_2012,Esterlis_2103}. In this limit, a controlled calculation of the fermion and boson self-energies is possible, and the model manifestly preserves the $\mathrm{LU}(1)$ conservation laws\footnote{This is contrast to some other approaches such as the dimensional regularization of Ref.~\cite{Dalidovich_1307}, which violates these conservation laws and hence risks disrupting the hydrodynamics.} so it is possible to make a direct comparison between the QBE and the hydrodynamic equations (see also the numerical results of Ref.~\cite{Wang_2209}). 
In particular, from the QBE one obtains \cite{Mandal_2108} \footnote{The calculations in Ref.~\cite{Mandal_2108} were actually framed in terms of the dimensional regularization of Ref.~\cite{Dalidovich_1307}, but as this gives the same form of the fermion and boson self-energy as the random-flavor large-$N$ model, up to different values of constants, the results should be the same.} an equation similar to \eqnref{eq:lineq}; however an important difference is that the ``Landau parameters'' in this equation have a non-trivial frequency dependence, while in our equation of motion the Landau parameters are defined in terms of a static susceptibility and have no frequency dependence. Thus, we need to examine this issue more closely to determine whether there is an inconsistency with the hydrodynamic result.

In Appendix \ref{appendix:qbe_solution}, we analyze the solution of the QBE. (Our conclusions are similar to those of Ref.~\cite{Kim_9504} which analyzed essentially the same equation, though we go into a bit more detail.) We show that if $n(\theta)$ varies sufficiently smoothly with $\theta$, then these equations of motion reduce to the hydrodynamic equations of motion \eqnref{eq:lineq}, with the effective Landau interactions $F(\theta,\theta')$ being zero\footnote{In interpreting this statement, the reader should keep in mind the terminological point made at the end of Section \ref{sec:ersatz_dynamics} regarding our definition of ``Landau interactions''.}. As a result, there is no zero sound mode, for instance. (While Ref.~\cite{Mandal_2108} claimed to obtain a zero sound mode, a more careful analysis of the solutions to their equations does not support their claim -- see Appendix \ref{appendix:qbe_solution}). 

There are two points to be made about this. Firstly, the calculations of Ref.~\cite{Mandal_2108} are computed with respect to a ``patch'' action that, while it is believed to capture the leading singularities associated with the critical fluctuations, is not necessarily expected to capture all of the IR properties of the system. Thus, the more appropriate interpretation of $F(\theta,\theta')$ being zero for this action is that in an actual microscopic system, the $F(\theta,\theta')$ in the IR theory would be nonzero but does not acquire any singular behavior due to the critical fluctuations. However, even this is probably not the physically correct statement. An essential feature of the quantum critical point is \cite{Shi_2204,Shi_2208} that $\xi$, given by \eqnref{eq:xi}, has a zero mode in the sense that there is a non-trivial function $u(\theta)$ such that $\int \xi(\theta,\theta') u(\theta) u(\theta') d\theta d\theta' = 0$. Since $\xi$ can be thought of as the inverse of the (infinite-dimensional) susceptibility matrix of the $N(\theta)$'s, this is equivalent to saying that susceptibility diverges in some channel, reflecting the fact that the $N(\theta)$'s mix with the order parameter at the critical point. These properties cannot be satisfied unless there is a relation between $F(\theta,\theta')$ and $v_F(\theta)$ enforced by the criticality. Thus, we feel that there is something that the QBE calculations as currently formulated are missing. We leave it for future work to ascertain which of the assumptions and approximations involved in the QBE calculations breaks down. Since, as we shall see, the QBE equations of motion do seem to give qualitatively reasonable results for modes where $n(\theta)$ is sharply peaked near some point of the Fermi surface, our suspicion is that there may be some subtle correlation between different patches on the Fermi surface, mediated by the boson, that is being missed.

Let us now return to analyzing the results obtained from the QBE equations of motion. One kind of mode that results are the so-called ``rough'' modes for which $n(\theta)$ becomes singular at particular points on the Fermi surface. (Another way to say this is that the Fourier-transformed quantities $n_l = \frac{1}{2\pi} \int e^{-il\theta} n(\theta) d\theta$, labelled by the angular momentum $l$, are not localized in angular momentum space but rather have a plane wave structure as $l\to \pm \infty$).
These exist as a continuum when $\omega < C q^{3/2}$, where $C$ is a constant. Such modes are not predicted by the zero-th order hydrodynamics, and going to higher orders in the derivative expansion presumably cannot give a fractional power in the dispersion relation. Thus, we are forced to conclude that the rough modes are completely invisible to hydrodynamics.

There is a simple physical picture for why one should not have expected these rough modes to be captured by hydrodynamics. The starting point of hydrodynamics is the assumption that the $N(\theta)$'s are conserved quantities. However, at nonzero frequency it is not strictly the case that $N(\theta)$ is conserved for each $\theta$, because the boson can scatter fermions between nearby points of the Fermi surface that are separated by less than $\Delta \theta_{\mathrm{crit}}(\omega) \propto \omega^{\alpha}$ [where $\alpha$ is some positive scaling exponent]. In other words, if we assume $\theta$ ranges over $[0,2\pi]$ and define the Fourier transformed quantities $N_l = \frac{1}{2\pi} \int e^{-il\theta} N(\theta) d\theta$, $N_l$ is approximately conserved only when $|l| \ll \Delta \theta_{\mathrm{crit}}(\omega)^{-1}$. However, the fact that $\Delta \theta(\omega) \to 0$ as $\omega \to 0$ means that the $N_l$'s are at least \emph{emergent} conserved quantities in the sense that for any fixed $l$, the relaxation rate of $n_l$ goes to zero as $\omega \to 0$. The problem for hydrodynamics is that with regard to the relaxation rate, the limits $l \to \infty$ and $\omega \to 0$ do not commute. Because the rough modes extend infinitely far in the $l$ space, they are very sensitive to this issue, while ``smooth'' modes that vary over the Fermi surface only on a scale $\gg \Delta \theta_{\mathrm{crit}}(\omega)$ should not be sensitive to this issue and thus are expected to be correctly described by hydrodynamics. (We emphasize that, since $\Delta \theta_{\mathrm{crit}}(\omega) \to 0$ as $\omega \to 0$, even very sharply peaked modes can still be described by hydrodynamics in the low-frequency limit.)

Thus, the mode spectrum predicted by zero-th order hydrodynamics and depicted in Figure \ref{fig:fl_spectrum} at the very least needs to be supplemented by adding in the non-hydrodynamic ``non-quasiparticle continuum'' for $\omega < C q^{3/2}$.
Let us discuss the fate of particle-hole spectrum as predicted from hydrodynamics. One might be concerned about the fact that according to the solution of the hydrodynamic equations of motion, the particle-hole spectrum involves modes that also appear to be ``rough''. One might have viewed this as hydrodynamics predicting its own breakdown.

In fact, however, this is not really the case, and we do expect on general grounds that the particle-hole continuum will survive. One argument for this is as follows. Suppose that we drive the system at frequency $\omega$ and wave-vector $q$, somewhere within the particle-hole spectrum region, with $q \sim v_F \omega$, but where the temporal driving is not strictly monochromatic, i.e.\ in frequency space there it has some small spread $\Delta \omega$. Then according to the hydrodynamic equations, the prediction is that one would excite a superposition of particle-hole continuum modes spread over a range $\Delta \theta \sim \Delta \omega/qv_F$ on the Fermi surface. So long as $\Delta \theta \gg \Delta \theta_{\mathrm{crit}}(\omega)$, this superposition will be well described by hydrodynamics. But, since $\Delta \theta_{\mathrm{crit}}(\omega) \to 0$ as $\omega \to 0$, it follows that if we hold $\omega/q$ fixed and take the limit as $\omega \to 0$, we can make the driving, and thus the modes that are excited, increasingly monochromatic as $\omega \to 0$ while still being well-captured by hydrodynamics. Thus, if for a given $q$, hydrodynamics breaks down for $\omega < \omega_{\mathrm{hydro}}(q)$ [for example, $\omega_{\mathrm{hydro}}(q) \sim q^{3/2}$ in the QBE result described above], it must be the case that $\omega_{\mathrm{hydro}}(q)/q \to 0$ as $q \to 0$.

Notwithstanding our concerns about the validity of the QBE calculations, one can verify that the above scenario is actually what occurs when solving the QBE equations of motion. More precisely, in Appendix \ref{appendix:qbe_solution} we find that for $\omega > \omega_{\mathrm{hydro}}(q)$,  the particle-hole continuum gets replaced by a ``pseudo-continuum'' comprising a set of discrete modes, but where the dispersion relations of neighboring modes have a relative spacing $\delta = \Delta \omega / \omega$ that is proportional to $\omega^{1/6}$ and hence goes to zero as $\omega \to 0$. The modes in this region are smeared out by an amount $\sim \delta$ over the Fermi surface.

\begin{figure}
    %
    %
   



    %
	\includegraphics{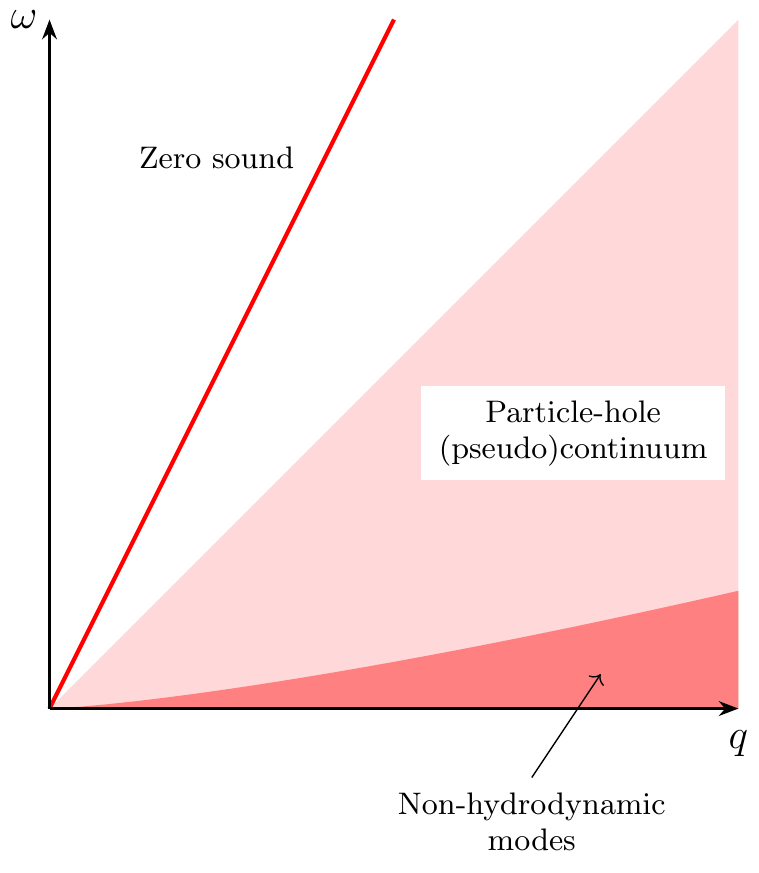}
        \caption{\label{fig:nonfl_spectrum} The expected features of the  mode spectrum for a general ersatz Fermi liquid without quasiparticles (the zero sound mode will be present if and only if it is predicted to exist from the zero-th order hydrodynamics.)}
    \end{figure}

Let us summarize the lessons that we have learned for a general ersatz Fermi liquid, going beyond the particular  models that we have been discussing in this section. By the definition of emergent symmetry, it must be the case that for a fixed $l$, the relaxation rate of $N_l$ goes to zero as $\omega \to 0$, but it is certainly possible that the limits $l \to \infty$ and $\omega \to 0$ do not commute. Nevertheless, the general arguments described above suggest that any breakdown of hydrodynamics can only happen for $\omega < \omega_{\mathrm{hydro}}(q)$, with $\omega_{\mathrm{hydro}}(q)/q \to 0$ as $q \to 0$. Outside of this region, hydrodynamics will remain valid (modulo the subtle issues about the particle-hole ``pseudo-continuum''). The general features expected of the mode spectrum are depicted in Figure \ref{fig:nonfl_spectrum}.

\section{Effect of quantum criticality on the dynamics}
\label{sec:effect_of_criticality}
Even though the zero-th order hydrodynamics of ersatz Fermi liquids is described by the same equation of motion as Fermi liquids, there is still one sense in which quantum critical points could conceivably exhibit qualitatively different dynamics compared to Fermi liquids. This is because a Fermi liquid always satisfies the property that $\xi(\theta,\theta')$ (thought of as an infinite-dimensional matrix) is positive-definite, as this is equivalent to the condition for absence of a Pomeranchuk instability. More generally, in any ersatz Fermi liquid, thermodynamic stability requires that $\xi(\theta,\theta')$ is always positive-\emph{semidefinite}. But at a quantum critical point one generally expects $\xi$ to support a non-trivial zero mode (for example see the computations of the susceptibility in Refs.~\cite{Shi_2204,Shi_2208}), i.e.\ it is positive-semidefinite but not positive-definite. One can ask how this affects the dynamics. 

Let us for simplicity first consider the case where the system is rotationally invariant. Then defining the Fourier coefficients $\xi(\theta,\theta') = \sum_l \xi_l e^{il(\theta - \theta')}$ and $n(\theta) = \sum_l n_l e^{i l \theta}$, and writing $\mathbf{k}_F(\theta) = k_F(\cos \theta, \sin \theta)$, \eqnref{eq:general_motion} (setting $\mathbf{E}=0$) becomes
\begin{equation}
    \label{eq:n_l_hopping}
    \omega n_l + \frac{mqk_F}{2(2\pi)^d} (\xi_{l-1} n_{l-1} + \xi_{l+1} n_{l+1}) = 0.
\end{equation}
Here we have taken the Fourier transform in space and time, introducing frequency $\omega$ and wave-vector $\mathbf{q} = (q,0)$. It is convenient to rewrite this in terms of the chemical potential $\mu_l = \xi_l n_l$, which gives
\begin{equation}
    \label{eq:xi_l_hopping}
    \omega \xi_l^{-1} \mu_l + \frac{mqk_F}{2(2\pi)^d} (\mu_{l-1} + \mu_{l+1}) = 0.
\end{equation}

This takes the form of a Schr\"odinger equation for a particle hopping on an infinite 1-D lattice with on-site potential $\propto \xi_l^{-1}$. Hence, in particular, if there is quantum criticality in a particular angular momentum channel $\pm l_*$ so that $\xi_{l_*} = \xi_{-l_*} = 0$, then we see that the on-site potential on sites $\pm l_*$ is \emph{infinite}.
Thus, we see that there are three different regions between which the particle cannot tunnel: $l > l_*$, $-l_* < l < l_*$, and $l < -l_*$. This implies, for example, that the particle-hole (pseudo)continuum will consist entirely of electrically neutral states since these modes (which correspond to unbound states which do not decay as $l \to \infty$) cannot penetrate the interior region, and hence have $n_0 = 0$. If $l_*=1$, we also find that there is an electrically charged mode which does not propagate, i.e.\ $\omega = 0$ regardless of $q$ (of course in general this mode will presumably have diffusive dynamics $\omega \propto iq^2$ once one includes higher-order terms in the hydrodynamic expansion). This can be regarded as closely related to the physics of ``critical drag'' \cite{Else_2010,Else_2106} which suppresses the DC conductivity that would otherwise be infinite.

It is worth asking whether these properties are robust to anisotropy. Among other things, we can think of anisotropy as introducing beyond-nearest neighbor hopping in \eqnref{eq:xi_l_hopping}. This allows the infinite potential barrier to be bypassed. Therefore, for generic $\xi(\theta,\theta')$ we expect that many of the peculiar properties discussed above will disappear in anisotropic systems and the dynamics will be qualitatively similar to a Fermi liquid. For example, as pointed out in Ref.~\cite{Shi_2208}, for an anisotropic system critical drag does not suppress the infinite DC conductivity without fine-tuning $\xi(\theta,\theta')$ [whether there might be reasons in a particular model why $\xi(\theta,\theta')$ would take special values, we leave as an open question].

\section{Extensions}
\label{sec:extensions}

\subsection{Spinful systems}
\label{subsec:spinful}
So far we have focused on spinless metals. Let us discuss extensions to the spinful case. As pointed out in Ref.~\cite{Else_2007} the emergent symmetry group of a spinful Fermi liquid (without spin-orbit coupling) is the group $\mathrm{LU}(1)_{\mathrm{spin}}$ defined by taking the quotient of $\mathrm{U}(2) \times \mathrm{LU}(1)$ by the diagonal $\mathrm{U}(1)$ subgroup. This has the same Lie algebra (though a different global structure) as $\mathrm{SU}(2) \times \mathrm{LU}(1)$, which reflects the fact that the \emph{charge} at each point on the Fermi surface is conserved, but only the \emph{total} spin is conserved, not the spin at each point on the Fermi surface.

One can therefore define a spinful ersatz Fermi liquid to be a non-Fermi liquid with the same emergent symmetry group $\mathrm{LU}(1)_{\mathrm{spin}}$ and attempt to study its hydrodynamics. In general, hydrodynamics of non-commuting conserved quantities is a much more challenging topic compared to Abelian hydrodynamics \cite{Leurs_0705}. However, for the zero-th order hydrodynamics we are considering in this paper the situation remains straightforward. The generators of the emergent symmetry group are the three components of the spin, which we write as a vector $\vec{S}$, and the charge $N(\theta)$ at each point on the Fermi surface. We can introduce the thermodynamically conjugate variables $\vec{h}$ and $\mu(\theta)$ respectively, such that a thermal equilibrium state is can be written as
\begin{equation}
    \frac{1}{\mathcal{Z}} \exp\left(-\beta\left[H - \int \mu(\theta) N(\theta) d\theta - \vec{h} \cdot \vec{S}\right]\right).
\end{equation}
The spin does not have any mixed anomaly with $N(\theta)$. Therefore, the spin falls into the category of ``spectator'' conserved quantity that we discussed in Section \ref{sec:zeroth_hydro}. For spectator conserved quantities it does not matter that they do not commute, and accordingly,
we find that the zero-th order hydrodynamics of the Fermi surface charges is again given by the identical form as in Section \ref{sec:ersatz_dynamics}.

We can compare this with Fermi liquid theory, where the dynamics is \emph{not} the same in the spinful case as in the spinless case, and in particular in the spinful case it is expressed \cite{Baym} in terms of the charge and spin at each point of the Fermi surface [or equivalently a $2\times2$ matrix-valued $N^{\alpha \beta}(\theta)$], despite the fact that the spin at each point on the Fermi surface is not a conserved quantity and therefore ought not to appear in hydrodynamics. Thus, unlike in the case of spinless Fermi liquid theory, spinful Fermi liquid cannot be interpreted as a hydrodynamic theory in the usual sense where only the densities of conserved quantities enter, though there could perhaps be some way to generalize the concept of hydrodynamics to accommodate it.

Going beyond Fermi liquids, in spinful non-Fermi liquids it is possible that the strict hydrodynamic description is recovered, giving dynamics different from spinful Fermi liquids. However, it is also possible that strict hydrodynamics will continue to break down and the dynamics could be more like that of a spinful Fermi liquid. We leave the exploration of these issues for future work.

    \subsection{Magnetic fields}
In (spinless) Fermi liquid theory, in the presence of a weak magnetic field there is a flow of quasiparticles not just in the spatial directions, but also in momentum space along the Fermi surface. Thus, restricting for simplicity to two spatial dimensions one is led to introduce also a current $j^\theta$ along the Fermi surface. The conservation equation \ref{eq:conservation_eqn}  gets generalized to
\begin{equation}
    \partial_t n(\theta) + \partial_i j^i(\theta) + \partial_\theta j^\theta(\theta) = \frac{m}{(2\pi)^d} \mathbf{w}(\theta) \cdot \mathbf{E}.
\end{equation}
In some ways we can think of the $\theta$ direction as being analogous to an extra spatial dimension. In particular, in Fermi liquid theory one can show that the result \eqnref{eq:extended_bloch_j} for the current can be extended to
\begin{equation}
    \label{eq:extended_bloch_j}
    j^a = \mu(\theta) \epsilon^{abc} \partial_b A_c,
\end{equation}
where the indices vary over the three dimensions $x,y,\theta$, and we have defined the vector potential $\mathcal{A}_c(\mathbf{x},t,\theta)$ in the following way: $\mathcal{A}_i = (k_F)_i(\theta) + A_i$ for $i=x,y$, where $A_i$ is the external electromagnetic vector potential. Formally one would expect that $\mathcal{A}_\theta$ represents the quasiparticle Berry connection on the Fermi surface, but we are assuming that this is independent of $x$ and $y$ and therefore does not enter \eqnref{eq:extended_bloch_j}.

In particular, the $\theta$ component of \eqnref{eq:extended_bloch_j} is $j^\theta = \mu(\theta) B$, where $B = \epsilon^{ij} \partial_i A_j$ is the magnetic field, while the spatial component gives $\mathbf{j}(\theta) = \mu(\theta) \mathbf{w}(\theta)$ as before. From this one can recover the usual collisionless Boltzmann equation for quasiparticles in a Fermi liquid moving in a magnetic field.

Does \eqnref{eq:extended_bloch_j} also hold in a general ersatz Fermi liquid in an equilibrium state? Observe that \eqnref{eq:extended_bloch_j} is precisely what one would get from applying the Bloch's theorem argument discussed above if one naively treats the $\theta$ direction as an extra spatial dimension.  Unfortunately, however, the $\theta$ dimension is not \emph{really} an extra spatial dimension; in particular, the $\theta$ dimension is compact and its size cannot be sent to infinity. Therefore, the Bloch's theorem arguments do not strictly apply, and we do not know how to prove that \eqnref{eq:extended_bloch_j} holds in a general ersatz Fermi liquid. Nevertheless, it seems to be a plausible conjecture. If this conjecture holds, it would imply that any ersatz Fermi liquid has the same dynamics as a Fermi liquid in zero-th order hydrodynamics, even in the presence of a weak magnetic field\footnote{Note that in linearizing the hydrodynamic equations of motion, we do not formally treat the magnetic field as being of the order of the perturbation, unlike the electric field. Otherwise \eqnref{eq:extended_bloch_j} would satisfy $\partial_\theta j^\theta = 0$ to linear order and not contribute to dynamics, assuming that $\mu(\theta)$ is independent of $\theta$ in the unperturbed state. Of course, as in Fermi liquid theory, the Boltzmann equation will presumably only be valid when the magnetic field is sufficiently small.}.

\section{Discussion}
\label{sec:discussion}
\subsection{Regime of validity for hydrodynamics}
\label{subsec:regime}
Since our results have been based on hydrodynamics, we should consider when exactly one should expect hydrodynamics to work. First of all, since we are assuming the quantities $N(\theta)$ are conserved, it is necessary to be at sufficiently low frequencies and temperatures, otherwise the conservation law can be broken by irrelevant operators. (Of course, as we saw in Section \ref{sec:qbe}, even in the limit as frequency and temperature go to zero, there can be some subtle issues of order of limits leading to the possibility of non-hydrodynamic ``rough'' modes). Therefore, if there are interesting non-Fermi liquid regimes at intermediate temperatures and frequencies then their dynamics \cite{Wang_2209} is not necessarily accessible using the techniques of this paper. In the remainder of this section, we will assume that we are in a regime where we can take the conservation laws to hold, and consider the conditions for hydrodynamics to be a valid description of the dynamics of the densities of the conserved quantities.

Firstly, we note that for temperature $T > 0$, there will likely be a thermalization timescale $\tau_{\mathrm{th}}(T)$. Generally in strongly coupled systems one expects $\tau_{\mathrm{th}}(T) \sim T^{-1}$. For frequencies $\omega \ll \tau_{\mathrm{th}}(T)^{-1}$, one certainly expects hydrodynamics to be valid. There is reason to suspect, however, that our hydrodynamic equations may in fact hold more generally than this. For one thing, in Fermi liquid theory, one can show that the quasiparticle distribution function actually remains thermal even for frequencies $\omega \gg T$, provided that $\omega$ and $T$ remain sufficiently small. This explains why the collisionless Boltzmann equation still holds in Fermi liquid theory even at $T=0$.

Beyond Fermi liquid theory, it is harder to make any general arguments. However, it is instructive to consider the results on metallic quantum critical points in Refs.~\cite{Shi_2204,Shi_2208}. There, the optical conductivity $\sigma(\omega)$ [at $\mathbf{q}=0$], was computed at $T=0$. It was found that
\begin{equation}
    \label{eq:resigma}
    \operatorname{Re} \sigma(\omega) = \mathcal{D} \delta(\omega) + \sigma_{\mathrm{inc}}(\omega),
\end{equation}
where the Drude weight $\mathcal{D}$ can be related to the susceptibilities of the $N(\theta)$'s. The first term is precisely what one would find as the prediction of the zero-th order hydrodynamics of this paper. The second term describes the corrections to zeroth-order hydrodynamics. Ref.~\cite{Shi_2204}  showed that in the model with a single-component fermion, $\sigma_{\mathrm{inc}}(\omega)$ is actually zero (up to the effects of irrelevant operators that were discarded). Thus, the hydrodynamic prediction is exactly correct in this case, even though $\omega \ll T$ is not satisfied. Meanwhile, in the ``random-flavor large $N$'' model considered in Ref.~\cite{Shi_2208}, it was found that $\sigma_{\mathrm{inc}}(\omega) \sim \omega^{-2/3}$. Such a fractional power presumably cannot be captured by hydrodynamics even going beyond zero-th order in the gradient expansion. However, as $\omega \to 0$ one can still think of this result as subleading compared to the hydrodynamic term (since if we pass to the full complex conductivity rather than just the real part, the hydrodynamic term will scale like $\sim i/\omega$, which diverges more rapidly than $\omega^{-2/3}$ as $\omega \to 0$).

\subsection{Properties \emph{not} constrained by our arguments}
What we have seen in this paper is that Fermi liquids and ersatz Fermi liquids in fact have many similarities in their dynamics. Let us, however, highlight areas where, notwithstanding our results, there remains a possibility for a difference between Fermi liquid and non-Fermi liquid behavior, or between different non-Fermi liquids.
One of them, of course, is the non-hydrodynamic modes that we discussed in Section \ref{sec:qbe}, as well as the possible corrections to hydrodynamics for $\omega \lesssim T$ discussed in Section \ref{subsec:regime}.
Another potential difference is in quantities related to the fermion Green's function, as measured for example in photoemission. This is not a hydrodynamic probe and is not constrained by the results discussed here. Moreover, at nonzero temperature or frequency, the rate at which the conservation law of the $N(\theta)$'s will be violated due to irrelevant operators can be different between Fermi liquids and non-Fermi liquids.

We also note that the constraints we have derived have only applied to the zero-th order hydrodynamics. To illustrate this limitation, consider the optical conductivity $\sigma(\omega)$, at $\mathbf{q}=0$.
In general one expects this to have the form \ref{eq:resigma}, where the delta function part can be derived from zero-th order hydrodynamics. Meanwhile, $\sigma_{\mathrm{inc}}(\omega)$ could potentially have a hydrodynamic description, at least for $\omega \ll T$, but it would require going to the next order in the hydrodynamic expansion. The first-order hydrodynamics is probably not constrained purely from the emergent symmetry and anomaly in the way that the zeroth-order hydrodynamics is.

Finally, we note that we have only considered the linearized equations of motion. It is an interesting question for the future to determine whether there are any statements that can be made about non-linear dynamics. Along these lines, see Ref.~\cite{Delacretaz_2203} for an intriguing perspective on the non-linear dynamics in the case of Fermi liquid theory, in which it was argued that the loop-group anomaly of Ref.~\cite{Else_2007} that we leveraged in the current work can be viewed as a linearized approximation to a more general structure.

\subsection{Relation with Refs.~\cite{Shi_2204,Shi_2208}}
Here we want to add a note of clarification regarding the relation between our results and those of Refs.~\cite{Shi_2204,Shi_2208}. In both cases, the results were presented roughly as arising ``due to the emergent symmetries and anomalies''. We wish to emphasize, however, that the arguments of Refs.~\cite{Shi_2204,Shi_2208}, though they invoked emergent symmetries, were still tied to the specific Hertz-Millis type models under consideration. Therefore, the results of Refs.~\cite{Shi_2204,Shi_2208} were much less general than those of the current work, which apply to any ersatz Fermi liquid. On the other hand, Refs.~\cite{Shi_2204,Shi_2208} also determined quantities such the incoherent part of the conductivity in \eqnref{eq:resigma}, which are not captured by the zero-th order hydrodynamics of this paper.

\subsection{Comparison with other approaches to metallic transport}
It is worth contrasting our results with those obtained from other approaches. Specifically, most previous works have not taken into account the emergent $\mathrm{LU}(1)$ symmetry and associated conservation laws. For example, the memory-matrix description of magnetotransport in Ref.~\cite{Lucas_1502} only took into account approximate conservation of energy and momentum. Such approaches may be valid in some regime of frequencies and temperatures, assuming that there is some separation between the timescale at which the $\mathrm{LU}(1)$ charges relax and the timescale at which momentum relaxes. But since the present paper is looking at the regime in which we can treat the $\mathrm{LU}(1)$ charges as conserved, our results are not directly comparable.

Similarly, there has been considerable interest in understanding transport in strongly coupled metals from the perspective of holography \cite{Holographic}, where the system is viewed as being dual to a weakly coupled gravitational theory in one higher dimension. The problem is that such models never seem to exhibit any sign of an emergent $\mathrm{LU}(1)$ symmetry. In our opinion, this should be viewed as a pathology of such models, given the general arguments for why an emergent $\mathrm{LU}(1)$ symmetry should be a generic feature of a compressible metal \cite{Else_2007}. In any case, this issue means that results from holography are not directly comparable to our results.

\subsection{Outlook: when does an emergent symmetry and its anomaly lead to dynamical modes?}
The fact that we get non-trivial dynamical modes from zero-th order hydrodynamics applied to the conserved quantities $N(\theta)$ is striking, because more commonly zero-th order hydrodynamics for conserved quantities arising from \emph{internal} symmetries is just trivial. (The situation is different for systems with momentum conservation, which arises from the non-internal continuous translation symmetry; for example one can think of the Euler equations of fluid dynamics as arising from zero-th order hydrodynamics.) The reason is that normally there will be some Bloch's theorem-like argument implying that the corresponding current density is zero in equilibrium, and therefore in zero-th order hydrodynamics. If all the currents are zero then the conservation equations just tell us that the local density of the conserved quantities is time-independent, i.e.\ we just have an equilibrium state.

As we saw, the reason why this does not apply for ersatz Fermi liquids is that the emergent symmetry has a 't Hooft anomaly, leading to a loophole in Bloch's theorem. Furthermore, in this case the anomaly actually dictates the zero-th order form of the constitutive relation for the current, so one ends up with dynamical modes in zero-th order hydrodynamics whose equations of motion are completely fixed (up to some thermodynamic susceptibility parameters) by the emergent symmetry and anomaly. This is precisely the same mechanism that is at work in the superfluid case \cite{Delacretaz_1908}.

We emphasize, however, that it is not the case that \emph{any} emergent symmetry and 't Hooft anomaly will lead to non-trivial zero-th order hydrodynamics. For example, a Weyl semimetal in 3 spatial dimensions has an emergent $\mathrm{U}(1) \times \mathrm{U}(1)$ symmetry corresponding to conservation of charge at each Weyl point. These emergent $\mathrm{U}(1)$ symmetries have a 't Hooft anomaly. However, there is no loophole to Bloch's theorem in this case, and the currents are always zero in an equilibrium state.

An interesting question for future work will be to determine whether there are any other systems where an emergent symmetry and anomaly leads to a loophole in Bloch's theorem, and hence to non-trivial zero-th order hydrodynamics. It was conjectured in Ref.~\cite{Else_2106} that this will occur in any \emph{compressible} system; that is, a system with microscopic lattice translation symmetry and charge conservation symmetry, such that the microscopic charge per unit cell, the filling $\nu$, can be continuously tuned. Ersatz Fermi liquids (and variations thereof) and superfluids are the main classes of compressible systems currently known: a more obscure case is the ``Bose-Luttinger liquid'' discussed in Refs.~\cite{Sur_1803,Lake_2101} (see Ref.~\cite{Else_2106} for the complete identification of the emergent symmetries and anomalies of the Bose-Luttinger liquid). An important open question is whether there are any fundamentally different possibilities.

In any case, let us note the following variations on ersatz Fermi liquids to which our arguments either apply directly, or could probably be extended. One example is the so-called $\mathrm{FL}^*$ \cite{Senthil_0209}, in which a Fermi liquid co-exists with a discrete topological sector leading to a violation of Luttinger's theorem. In this case, in addition to the emergent $\mathrm{LU}(1)$ symmetry, there is also a finite 1-form symmetry. However, finite symmetries do not enter hydrodynamics, so our results will carry over directly to this case. A more subtle example is a Fermi surface built from fractionally charged particles (which are allowed if there is also a topological sector). The precise nature of the emergent symmetry group in this case has not been spelled out in the literature, which would be a minimal prerequisite to extending our results to this case. A final case that can be considered is that of a Fermi surface coupled to a fluctuating gauge field, as occurs in the case of spinon Fermi surfaces, or in the composite Fermi liquids of the fractional quantum Hall effect \cite{Halperin__1993,Son_1502}. In this case the symmetry group was argued in Ref.~\cite{Else_2007} to be a non-Abelian central extension of $\mathrm{LU}(1)$. The non-commutativity of the conserved quantities may make constructing the zero-th order hydrodynamics in this case a more challenging task, but it would be interesting to see if it leads to any differences in the dynamics.

\begin{acknowledgments}
    I thank T.~Senthil, Zhengyan Darius Shi, and Luca Delacrétaz for helpful discussions. I was partly supported by the EPiQS initiative of the Gordon and Betty Moore foundation, grant nos.~GBMF8683 and GBMF8684. Research at Perimeter Institute is supported in part by the Government of Canada through the Department of Innovation, Science and Economic Development and by the Province of Ontario through the Ministry of Colleges and Universities.
\end{acknowledgments}

\appendix
\section{Susceptibilities in Fermi liquid theory}
\label{appendix:fl_susceptibility}
Here we will compute $\xi(\theta,\theta')$ in Fermi liquid theory. In Fermi liquid theory in equilibrium at inverse temperature $\beta$, the distribution function $f$ follows a Fermi-Dirac distribution
\begin{equation}
f(\mathbf{k}) = \frac{1}{(2\pi)^d} \frac{1}{1 + e^{\beta(\epsilon(\mathbf{k})-\mu)}}
\end{equation}
with $\epsilon(\mathbf{k})$ given by \eqnref{eq:qp_energy} [thus, this is actually an implicit equation for $f$]. If we apply a $\theta$-dependent chemical potential, we can generalize this to
\begin{equation}
    \label{eq:fk_muth}
    f(\mathbf{k}) = \frac{1}{(2\pi)^d} \frac{1}{1 + e^{\beta(\epsilon(\mathbf{k})-\mu(\theta_\mathbf{k}))}}
    \end{equation}
    where $\theta_\mathbf{k}$ describes the point on the Fermi surface to which $\mathbf{k}$ is closest. Now we want to introduce a perturbation $\mu(\theta) = \mu + \delta \mu(\theta)$ and compute the corresponding change $\delta n(\theta)$ at linear order, keeping the temperature $T$ fixed. To this end, we can approximate
    \begin{equation}
        \label{eq:eps_approx}
        \epsilon(\mathbf{k}) = \epsilon_0(\theta_\mathbf{k}) + \delta \epsilon(\theta_\mathbf{k}) + \mathbf{v}_F(\theta_\mathbf{k}) \cdot (\mathbf{k} - \mathbf{k}_F(\theta_\mathbf{k}))
    \end{equation}
    where we have dropped terms that are higher-order in $\mathbf{k} - \mathbf{k}_F(\theta)$ and in the perturbation, and we defined
    \begin{equation}
        \label{eq:deltaeps}
        \delta \epsilon(\theta) = \int F(\theta,\theta') \delta n(\theta') d\theta'.
    \end{equation}
    Substituting into \eqnref{eq:fk_muth}, we find the effect of the perturbation on the distribution function in the vicinity of any given $\theta_\mathbf{k}$ is only to translate it by an amount $\delta k_F(\theta_\mathbf{k}) \hat{\mathbf{w}}(\theta_\mathbf{k})$ in momentum space, where
    \begin{equation}
    \delta k_F(\theta) = \frac{1}{v_F(\theta)} [-\delta \epsilon(\theta) + \delta \mu(\theta)]. 
    \end{equation}
   It follows that
    \begin{equation}
        \delta n(\theta) = \frac{1}{(2\pi)^d v_F(\theta)} |\mathbf{w}(\theta)| [-\delta \epsilon(\theta) + \delta \mu(\theta)]
    \end{equation}
    Substituting \eqnref{eq:deltaeps} and rearranging we find
    \begin{equation}
        \delta \mu(\theta) = (2\pi)^d \frac{v_F(\theta)}{|\mathbf{w}(\theta)|} \delta n(\theta) + \int F(\theta,\theta') \delta n(\theta') d\theta'
    \end{equation}
    from which we can read off that
    \begin{equation}
        \label{eq:xi_T}
        \left[\frac{\partial \mu(\theta)}{\partial n(\theta')}\right]_T = (2\pi)^d \frac{v_F(\theta)}{|\mathbf{w}(\theta)|} \delta(\theta - \theta') + F(\theta,\theta').
    \end{equation}
    This almost gives \eqnref{eq:xi}; we just need to worry about the fact that \eqnref{eq:xi} involves a derivative at constant entropy rather than constant temperature. However, recall that in Fermi liquid theory the entropy density is proportional to
    \begin{equation}
        \label{eq:landau_S}
        \int \bigl \{ n(\mathbf{k})  \log n(\mathbf{k})  + [1-n(\mathbf{k})] \log[1 - n(\mathbf{k})] \bigr \} d^d \mathbf{k}
    \end{equation}
    where $n(\mathbf{k}) = (2\pi)^d f(\mathbf{k})$.
    Substituting \eqnref{eq:fk_muth} and \eqnref{eq:eps_approx} into \eqnref{eq:landau_S}, we find, using the observation about the distribution just getting shifted, that the entropy density does not change due to the perturbation at this order.
     Hence, the derivative at constant entropy is also given by \eqnref{eq:xi_T}.

\section{Energy current in the equilibrium state}
\label{sec:energy_currents}
Here we will consider a system with microscopic energy conservation and charge conservation, and an emergent $\mathrm{LU}(1)$ symmetry for which the microscopic $\mathrm{U}(1)$ embeds a subgroup,  i.e.\ the microscopic charge $Q_{UV}$ can be expressed in the low-energy theory as $\int N(\theta) d\theta$. We wish to determine the expectation value of the energy current in the thermal equilibrium state
\newcommand{\UV}{\mathrm{UV}}
\newcommand{\IR}{\mathrm{IR}}
\begin{equation}
    \label{eq:rho_mu}
    \rho = \frac{1}{\mathcal{Z}} \exp\left[-\beta\left(H_{\IR}- \int \mu(\theta) N(\theta) d\theta \right)\right]
\end{equation}
where $H_{{\mathrm{IR}}}$ is the Hamiltonian of the low-energy theory.
In Ref.~\cite{Kapustin_1904} it was proven that the energy current in a thermal equilibrium state of a lattice model is zero. The issue with applying this result in the present context is that, although we can assume that the low-energy Hamiltonian $H_{\mathrm{IR}}$ can emerge out of a microscopic lattice model with Hamiltonian $H_{\mathrm{UV}}$, the argument of Ref.~\cite{Kapustin_1904} only applies to thermal equilibrium states of the form $Z^{-1} e^{-\beta H_{\mathrm{UV}}}$, and hence by extension to $Z^{-1} e^{-\beta H_{\mathrm{IR}}}$, and not \eqnref{eq:rho_mu}. It would not be so bad if $\mu(\theta)$ were independent of $\theta$ because we could then simply apply the arguments of Ref.~\cite{Kapustin_1904} to the Hamiltonian $H_{\mathrm{UV}}-  \mu Q_{\mathrm{UV}}$. Otherwise, the situation is trickier.

Nevertheless, we expect that if $\mu(\theta)$ is at most a small perturbation on top of a $\theta$-independent $\mu$ (which is the situation for which we want to compute the dynamics in the current paper), it will be possible to deform the lattice Hamiltonian $H_{\mathrm{\UV}}$ to a different lattice Hamiltonian $H_{\UV}'$ for which $H_{\IR}' = H_{\IR} - \int [\mu(\theta) - \mu]$ captures the low-energy theory. For example, in a non-interacting Fermi gas, $H_{\IR}'$ just corresponds to shifting the dispersion relation near the Fermi surface by
 $\epsilon_{\mathbf{k}} \to \epsilon_{\mathbf{k}} + [\mu(\theta_\mathbf{k}) - \mu]$.
 If we define a smooth function $\delta \epsilon_{\mathbf{k}}$ in momentum space such that $\delta \epsilon_{\mathbf{k}} = \mu(\theta_\mathbf{k}) - \mu$ near the Fermi surface, then we can define $H_{\UV}' = H_{\UV} + \sum_{\mathbf{k}} \delta \epsilon_{\mathbf{k}} \psi_{\mathbf{k}}^{\dagger} \psi_{\mathbf{k}}$ (where the added term is local on the lattice with at most exponentially decaying tails by the smoothness of $\delta \epsilon_{\mathbf{k}}$).
 
 In general, if it is true that one can find such an $H_{\UV}'$,  then we can define $K_{\UV} = H_{\UV}' + \mu Q_{\UV}$, for which the corresponding low-energy Hamiltonian is
 \begin{equation}
 K_{\IR} = H_{\IR} - \int \mu(\theta) N(\theta) d\theta
 \end{equation}
  and apply the arguments of Ref.~\cite{Kapustin_1904} to $K_{\UV}$ to conclude that the energy current is zero. The one remaining issue is that, although $\rho = Z^{-1} e^{-\beta K_{\IR}}$ gives the same \emph{state} as \eqnref{eq:rho_mu}, the definition of the energy current operator depends on the Hamiltonian. Therefore, we need to take into account the relation between the energy current operators $j^{K_{\IR}}$ and $j^{H_{\IR}}$. It is clear that we should set
  \begin{equation}
    \mathbf{j}^{K_{\IR}} = \mathbf{j}^{H_{\IR}} - \int \mu(\theta) \mathbf{j}(\theta) d\theta,
  \end{equation}
  where $\mathbf{j}(\theta)$ is the current of $N(\theta)$. Therefore, if $\langle \mathbf{j}^{K_{\IR}} \rangle = 0$ in the state $\rho$, we conclude that
\begin{equation}
  \langle \mathbf{j}^{H_{\IR}} \rangle = \int \mu(\theta) \langle \mathbf{j}(\theta) \rangle d\theta.
\end{equation}

\section{Solving the equations of motion arising from the QBE}
\label{appendix:qbe_solution}
In this appendix we will solve the equations of motion derived in Ref.~\cite{Mandal_2108} from the QBE. For simplicity, we will focus on the rotationally-invariant case. Then the equations of motion are best expressed in terms of the coefficients of the Fourier series, defined by $n(\theta) = \sum_l n_l e^{il\theta}$.
The result of Ref.~\cite{Mandal_2108}, upon also taking the Fourier transform in space and time with frequency $\omega$ and wave-vector $q$, is:
\begin{equation}
    \label{eq:qbe_hopping}
    \omega  n_l = \frac{q v_F}{2} \frac{ n_{l-1} + n_{l+1} } { 1 + \widetilde{F}_0 - \widetilde{F}_l}`
\end{equation}
where $\widetilde{F}_l$ is the Fourier series of a function that can be approximated as

\begin{equation}
    \label{eq:Ftheta}
    \widetilde{F}(\theta) = \begin{dcases} \frac{1}{\pi k_F^2 \theta_{\mathrm{crit}}^2} & |\theta| < \theta_{\mathrm{crit}} \\ \frac{1}{\pi k_F^2 \theta^2} & |\theta| > \theta_{\mathrm{crit}}
    \end{dcases}
\end{equation}
with $\theta_{\mathrm{crit}} = (|\omega|/\omega_0) ^{1/3}$, and $\omega_0$ a constant.

Observe that this is not actually periodic in $\theta$. Ultimately this reflects the fact that $\widetilde{F}(\theta)$ was computed within the theory that is supposed to describe the dynamics of a pair of antipodal patches on the Fermi surface rather than the whole Fermi surface. However, for $\theta_{\mathrm{crit}} \ll 1$ this function is sharply peaked near $\theta=0$ and one can imagine that this singular behavior within a patch should contain the important physics. Formally we can then replace $\widetilde{F}(\theta) \to \widetilde{F}(\theta)^{\mathrm{periodic}} := \sum_{n=-\infty}^{\infty} \widetilde{F}(\theta + 2\pi n)$ to get a periodic function. The important point, however, is that this implies that the Fourier series of $\widetilde{F}^{\mathrm{periodic}}$, which formally is defined only for integer argument, can actually be lifted to a smooth function of a continuous argument, namely the continuous Fourier transform of the original $\widetilde{F}(\theta)$ given by \eqnref{eq:Ftheta}.

If $\theta_{\mathrm{crit}} \ll 1$, then for $|l| \ll 1/\theta_{\mathrm{crit}}$, we can approximate $\widetilde{F}_l$ by the first two terms of the Taylor series [keeping in mind that the inversion symmetry of $\widetilde{F}(\theta)$ implies that $\widetilde{F}_l = \widetilde{F}_{-l}$]:
\begin{equation}
    \label{eq:Fl_taylor}
    \widetilde{F}_l = \frac{1}{\theta_{\mathrm{crit}}} \left( c_0 - c_1 (\theta_{\mathrm{crit}} l)^2 + \cdots \right)
\end{equation}
where $c_0$ and $c_1$ are dimensionless constants of order 1.
In particular, we see that if $|l| \ll 1/\delta$, where $\delta := \sqrt{\theta_{\mathrm{crit}}}$, then $|\widetilde{F}_0 - \widetilde{F}_l| \ll 1$. So in that case we can approximate \eqnref{eq:qbe_hopping} by
\begin{equation}
    \omega n_l = \frac{q v_F}{2} (n_{l-1} + n_{l+1}),
\end{equation}
which agrees with \eqnref{eq:lineq} setting $F_l = 0$.

Next we will go beyond this approximation and find the solutions to \eqnref{eq:qbe_hopping}.
Defining $\Omega = \omega/(qv_F)$, \eqnref{eq:qbe_hopping} becomes
\begin{equation}
    n_l ( 1 + \widetilde{F}_0 - \widetilde{F}_l ) = \frac{1}{2 \Omega} (n_{l-1} + n_{l+1}).
\end{equation}
We can view this as an effective Schr\"odinger equation for a particle hopping on 1D lattice with potential $V_l =  1 + \widetilde{F}_0 - \widetilde{F}_l$. Bound states of this potential will correspond to discrete modes in the oscillation spectrum, while unbound states will correspond to a continuum. Let us first observe that for $|l| \ll 1/\theta_{\mathrm{crit}}$, $V_l \approx 1$, while for $l \to \pm \infty$, $V_l \to 1 + \widetilde{F}_0 = 1 + c_0 \theta_{\mathrm{crit}}^{-1}$.

\textbf{Unbound states: non-quasiparticle continuum}. An unbound state will have the asymptotic form as $l \to \infty$:
\begin{equation}
    n_l = A_+ e^{i \Theta l} + B_+ e^{-i \Theta l}
\end{equation}
and as $l \to -\infty$:
\begin{equation}
    n_l = A_- e^{i \Theta l} + B_- e^{-i \Theta l}
\end{equation}
for some constants $A_\pm$, $B_\pm$, and where $\pm \Theta$ are the real solutions to $\cos \Theta = \Omega (1 + c_0 \theta_{\mathrm{crit}}^{-1})$. For any given $\Theta$, there will be some scattering matrix $\mathbb{S}(\Theta)$ such that the coefficients are required to obey
\begin{equation}
    \begin{bmatrix} A_+ \\ B_+ \end{bmatrix} = \mathbb{S}(\Theta) \begin{bmatrix} A_- \\ B_- \end{bmatrix}
\end{equation}
which will generically have solutions. Thus, we find that there are continuum modes whenever $\cos \Theta = \Omega(1 + c_0 \theta_{\mathrm{crit}}^{-1})$ has a solution for $\Theta$; or in other words whenever $|\Omega|(1 + c_0 \theta_{\mathrm{crit}}^{-1}) \leq 1$. This defines what in the main text we called the ``non-quasiparticle continuum'', and its boundary occurs at $|\omega|/(qv_F) = (1 + c_0 \theta_{\mathrm{crit}}^{-1})^{-1}$. As $\theta_{\mathrm{crit}} = (|\omega|/\omega_0)^{1/3}$. the asymptotic form of the boundary as $\omega \to 0$ scales like $|\omega| \propto q^{3/2}$.

\textbf{Bound states: pseudo-continuum}. For $\theta_{\mathrm{crit}} \ll 1$, $V_l$ defines a very flat potential well. Therefore, the ``bound states'' will actually be very spread out in the $l$ space. and moreover we can imagine solving the problem through a lattice version of the WKB approximation, in which the wavefunction is approximated locally near a given point $l_0$ as $n_l = A(l_0) \cos(\kappa(l_0) l +  \phi(l_0))$, where $A(l)$ and $\phi(l)$ are slowly varying functions of $l$, and $\kappa(l)$ satisfies
\begin{equation}
\cos \kappa(l) = \Omega V_l.
\end{equation}
Since $V_l \to 1$ as $l \to 0$, a necessary condition for a solution is that $|\Omega| < 1$.
The lattice equivalent of the ``classical turning points'', where the wavefunction crosses over from being oscillatory to exponentially decaying, occur at $l=\pm l_*$, where $|\Omega| V_{l_*} = 1$.

For our purposes, the only thing we will be interested in is the quantization condition that determines the discrete values of $\Omega$ for which a bound-state solution exists. We can determine this approximately by demanding that the lattice version of the semiclassical quantization condition be satisfied, namely:
\begin{equation}
    \label{eq:scquant}
\frac{1}{2\pi} \int_{-l_*}^{l_*} \cos^{-1} \left( \Omega V_l \right) dl = n + \frac{1}{2}.
\end{equation}
where $n$ is an integer.

For $\theta_{\mathrm{crit}} \ll 1$ and $\Omega \sim 1$, the particle will be confined to the region where $l \ll \theta_{\mathrm{crit}}^{-1}$. Hence we can again invoke the Taylor series \ref{eq:Fl_taylor}, but we now keep the quadratic term. Then the solution for the turning point can be written as
\begin{equation}
    l_* = \sqrt{c_1^{-1} \theta_{\mathrm{crit}}^{-1} (|\Omega|^{-1} - 1)}
\end{equation}

Making the change of variables $u = l/l_*$, \eqnref{eq:scquant} can then be rewritten as
\begin{equation}
    \label{eq:Omega_condn}
    \mathcal{I}(\Omega) = \sqrt{\theta_{\mathrm{crit}}}(n + 1/2)
\end{equation}
where we defined
\begin{equation}
    \mathcal{I}(\Omega) := \sqrt{c_1^{-1} (\Omega^{-1} - 1)} \int_{-1}^1 \cos^{-1} (\Omega - (\Omega - 1)u^2) du.
\end{equation}
This implies that for $\theta_{\mathrm{crit}} \ll 1$, the oscillation modes occur at $\Omega = \Omega_n$, where $\Omega_n$ is the solution to \eqnref{eq:Omega_condn}. The spacing between adjacent solutions for $\Omega$ is given by $\Delta \Omega \approx |\mathcal{I}'(\Omega)|^{-1} \sqrt{\theta_{\mathrm{crit}}}$, and in particular is proportional to $\delta := \sqrt{\theta_{\mathrm{crit}}}$ as stated in the main text.

Finally, we note that the spatial extent of the wavefunction in $l$ space is $\sim l_* \propto \theta_{\mathrm{crit}}^{-1/2}$. Hence, the mode occupies a region of width $\Delta \theta \propto \sqrt{\theta_{\mathrm{crit}}} = \delta$ on the Fermi surface.

\textbf{Edge of the particle-hole pseudo-continuum}. Near the upper edge of the particle-hole pseudocontinuum, where $\Omega \approx 1$, the WKB approximation will begin to break down. Instead, let us assume that $n_l$ varies slowly on the lattice scale as a function of $l$. Then we can Taylor expand $n_{l \pm 1} = n_l \pm \partial_l n_l + \frac{1}{2} \partial_l^2 n_l$ in \eqnref{eq:qbe_hopping}, giving
\begin{equation}
    n_l V_l \Omega = n_l + \frac{1}{2} \partial_l^2 n_l
\end{equation}
Let us furthermore assume that the particle is confined to the region where $l \ll \theta_{\mathrm{crit}}^{-1}$. Hence, again keeping terms up to quadratic order in $l$ in \eqnref{eq:Fl_taylor}, we obtain
\begin{equation}
     n_l \Omega c_1 \theta_{\mathrm{crit}} l^2 - \frac{1}{2} \partial_l^2 n_l = n_l(1 - \Omega)
\end{equation}
This can be interpreted as the Schr\"odinger equation (now in continuous space) for a harmonic oscillator. We conclude that it has solutions when
\begin{equation}
    1 - \Omega = \sqrt{2 c_1 \theta_{\mathrm{crit}} \Omega} \left(n + 1/2\right).
\end{equation}
for some integer $n \geq 0$.
When $\theta_{\mathrm{crit}} \ll 1$ and $n \sim 1$ we can approximate $\Omega \approx 1$ on the right-hand side, giving
\begin{equation}
    1 - \Omega = \sqrt{2 c_1 \theta_{\mathrm{crit}}} \left(n + 1/2\right).
\end{equation}
For $\theta_{\mathrm{crit}} \ll 1, n \lesssim 1, \Omega \approx 1$ we can verify that the assumptions we have made are self-consistent. Like the solutions deeper inside the particle-hole pseudo-continuum, these solutions are also spread out by an amount $\sim \delta$ on the Fermi surface. As $n$ becomes larger (and hence $\Omega$ moves away from 1), these solutions will transition into the WKB solutions found earlier. 

\textbf{Absence of zero sound}. Observe that none of the solutions that we have found correspond to zero sound, contrary to the claim of Ref.~\cite{Mandal_2108}. The issue is that Ref.~\cite{Mandal_2108} did not properly solve \eqnref{eq:qbe_hopping} and instead made an ``ansatz'' $n_l = e^{-\kappa |l|}$ for $|l| > 1$ that is not actually a solution because $V_l$ is not a constant function of $l$, except in the limit $\theta_{\mathrm{crit}} \to 0$.
If one takes this limit then $V_l \to 1$ and there is still no zero sound.

\bibliography{ref-autobib,ref-manual}
\end{document}